\documentclass[fleqn,12pt,twoside]{article}

\usepackage[headings]{espcrc1}

\usepackage{epsfig}
\usepackage{wrapfig}

\title{Azimuthal correlations of high-$p_{T}$ photons and hadrons in
  Au+Au collisions at RHIC}

\author{T. Dietel (for the STAR Collaboration\footnote{For the full
    list of STAR authors and acknowledgements, see appendix
    'Collaborations' of this volume.})
  \address{Johann-Wolfgang-Goethe Universit\"at,
    Institut f\"ur Kernphysik\\
    Max-von-Laue-Str. 1,
    60438 Frankfurt am Main,
    Germany}}

\runtitle{Azimuthal photon-hadron correlations in Au+Au collisions at
  RHIC}

\runauthor{T. Dietel}

\begin{document}

\maketitle

\begin{abstract}
  The measurement of $\gamma$+jet events with a direct photon in
  coincidence with an energetic parton can provide unique insights
  into the propagation and fragmentation of the parton in the presence
  of the hot and dense medium created in heavy-ion collisions. One way
  to explore these effects is the study of azimuthal correlations
  between the direct photon and hadrons produced in the fragmentation.

  We present azimuthal correlations between photons with a transverse
  momentum of more than 10\,GeV and charged hadrons with a transverse
  momentum of more than 2\,GeV in Au+Au collisions at $\sqrt{s_{NN}} =
  200\,\mbox{GeV}$ that have been measured with the STAR detector at
  RHIC. The separation of correlations with direct photons and photons
  from hadronic decays will be discussed.
\end{abstract}

\section{INTRODUCTION}

Photon-tagged jets in heavy ion collisions provide the ideal framework
to study the interaction of energetic partons traversing a strongly
interacting, hot and dense medium like the Quark-Gluon Plasma
\cite{harris_muller/QGP}.  Annihilation ($q\overline{q} \rightarrow
g\gamma$) or Compton scattering ($qg \rightarrow q\gamma$) results in
$\gamma+$jet events with a prompt photon, which does not interact
strongly, and a parton, which is subject to energy loss in the medium
and subsequent fragmentation into a jet of hadrons \cite{wang/eloss}.
As the photon escapes the medium without interactions, the analysis of
these events will provide the best possible determination of the
initial kinematics of the parton, and therefore all modifications of
the jet can be attributed to the interaction of the parton or the
emerging jet with the medium \cite{wang/jq_tagphoton,wang/gammajet}.
Azimuthal correlations with prompt photons as trigger particles
reflect these interactions and provide information about the energy
loss mechanism during the propagation of the parton through the medium
\cite{filimonov/gammajet}.

\section{ANALYSIS}

The data presented here has been taken by the STAR experiment during a
long Au+Au run at top energy of $\sqrt{s_{NN}} = 200\,\mbox{GeV}$ at
the RHIC collider. Events triggered on a high tower
($E_{T}>9\,\mbox{GeV}$) in the Barrel Electromagnetic Calorimeter
(BEMC) were selected for fast analysis. In the configuration used for
this analysis, about 180 million events were offered to the trigger,
corresponding to an integrated luminosity of $25\,\mu\mbox{b}^{-1}$.

\begin{wrapfigure}{r}{0.49\textwidth}

\epsfig{file=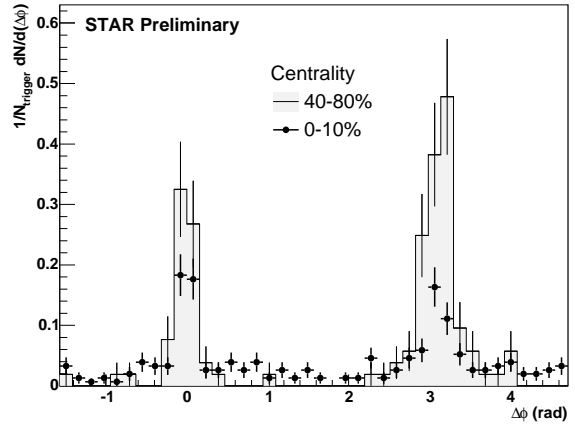, width=0.48\textwidth}

\caption{$\Delta\phi$ correlation of trigger BEMC clusters with
$E_{T}^{trigger} > 10\,\mbox{GeV}$ and associated charged particles
with $4\,\mbox{GeV} < p_{T}^{assoc} < E_{T}^{trigger}$. The
correlation function is corrected for tracking efficiency.
Combinatorial background has not been subtracted.}

\label{dphi_pc}

\end{wrapfigure}

This analysis studies azimuthal correlations between neutral clusters
in the BEMC \cite{nim/bemc} and charged tracks reconstructed with the
Time Projection Chamber (TPC) \cite{nim/tpc}.

Neutral trigger particles are reconstructed by adding the energy of
the three neighboring towers with the highest energy deposition to a
trigger tower. For each event, only the trigger cluster with the
largest transverse energy is used and only if $E_{T}>10\,\mbox{GeV}$.
The cluster position is defined as the center of the highest energy
tower of the cluster. Clusters with a matching TPC track with
$p_{T}>1\,\mbox{GeV}$ within 0.03 in $\eta$ and $\phi$ are vetoed to
remove charged particles with large energy deposition.

Charged particles with high transverse momenta are then associated
with these neutral trigger particles.  Due to large acceptance losses
near the TPC edges, the acceptance has been limited to $|\eta|<0.9$.

Figure \ref{dphi_pc} shows the distribution of azimuthal angles
between trigger clusters with $E_{T}=10-15\,\mbox{GeV}$, which are a
mixture of neutral pions and photons, and associated particles with
$4\,\mbox{GeV} < p_{T}^{assoc} < E_{T}^{trigger}$. A correction for
the tracking efficiency of the associated particle has been applied.
No background has been subtracted, and the reduced near-side yield due
to the charged track veto has not been corrected for.  Note that the
background is very low. This greatly reduces systematic uncertainties
on the extracted properties (both shape and yield) of the correlation
signal compared to previous analyses at lower $p_{T}$
\cite{star/highpt/b2b}.  Clear near- and away-side correlations are
visible in peripheral as well as central events, in contrast to the
disappearance of back-to-back correlations between charged particles
at lower transverse momenta in central collisions
\cite{star/highpt/b2b}.

The associated yield $n_{near}$, $n_{away}$ per trigger particle is
determined by integration over the near- ($|\Delta\phi|<0.3$) or
away-side peaks ($|\Delta\phi-\pi|<0.6$), subtraction of a flat
background determined from the region between the peaks, and
normalization with the number of trigger particles in the event
sample.

\begin{figure}
\begin{minipage}[t]{0.48\textwidth}

\epsfig{file=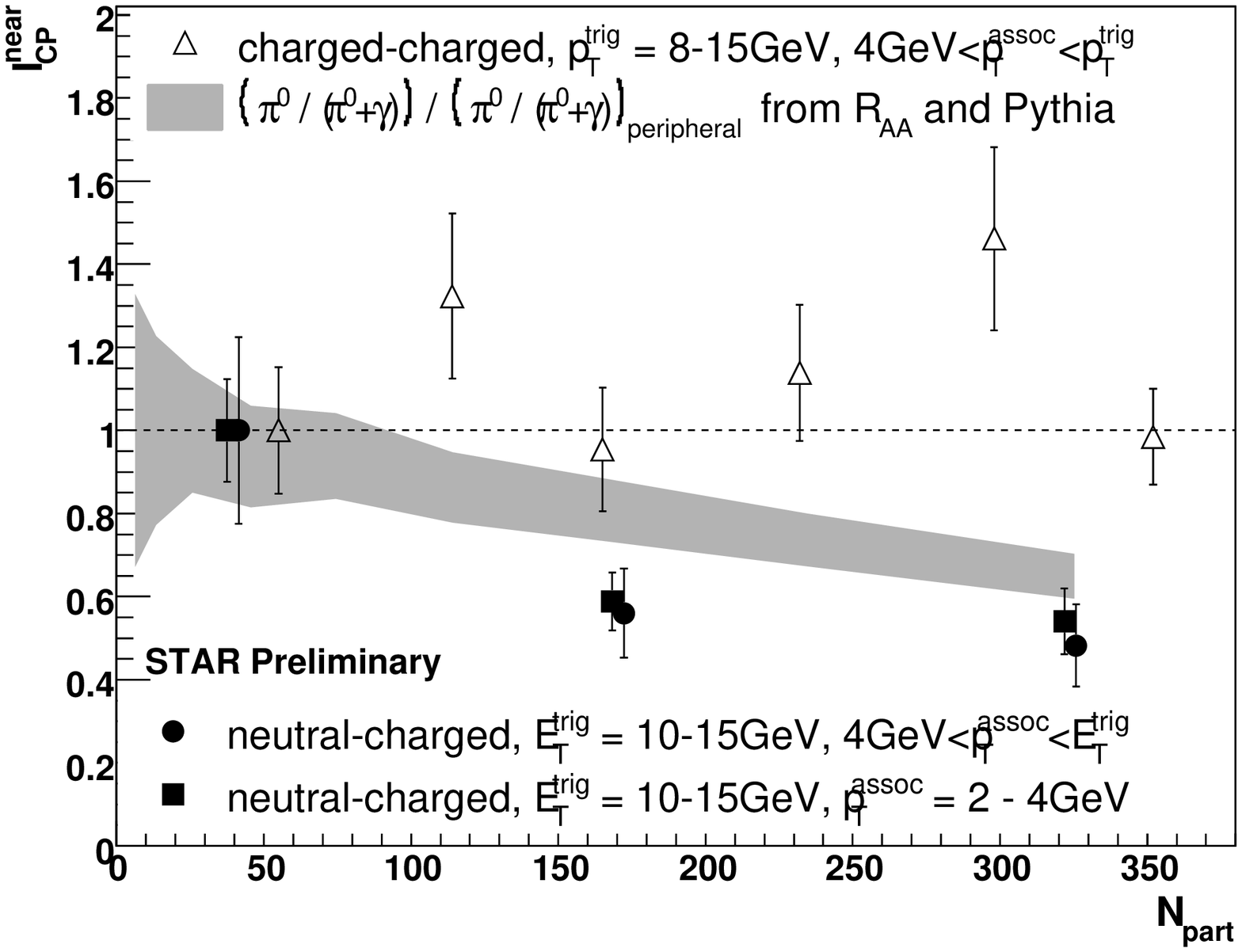, width=1.00\textwidth}

\caption{Ratio of the near-side yield in central and peripheral
  collisions $I_{CP}^{near}$ for neutral-charged (solid symbols) and
  charged-charged correlations (open symbols), and expected $\pi^0$
  fraction relative to peripheral collisions (grey band, see text).}

\label{icp_near}

\end{minipage}
\hfill
\begin{minipage}[t]{0.48\textwidth}
\epsfig{file=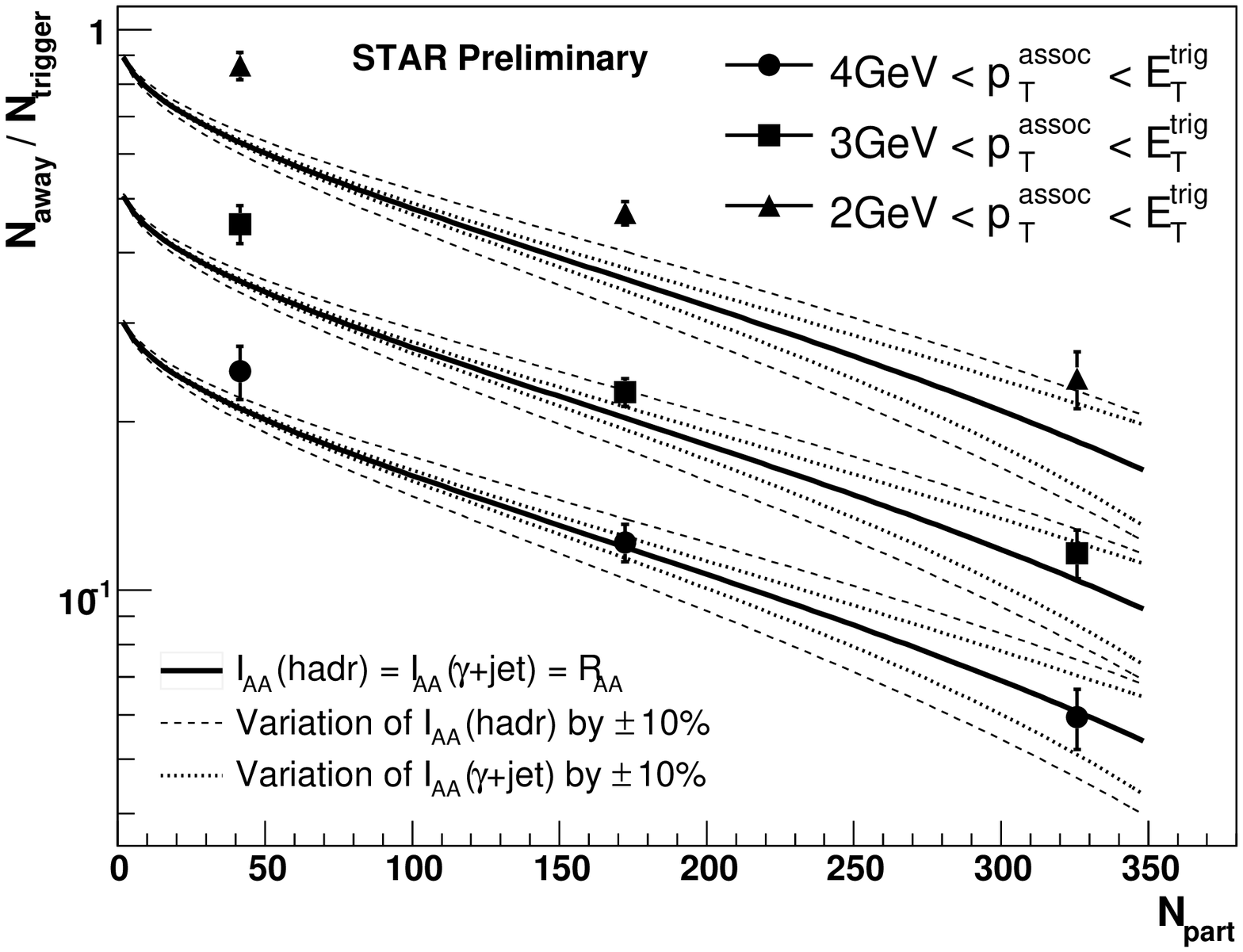, width=1.00\textwidth}

\caption{Away-side yield per trigger cluster for neutral-charged
  correlations. Solid lines are away-side yield from Pythia with a
  suppression of $\gamma+$jet and dijet components by $R_{AA}$
  ($I_{AA}(\mbox{hadr}) = I_{AA}(\gamma+\mbox{jet}) = R_{AA})$. The
  dashed and dotted lines represent 10\% variations of
  $I_{AA}(\mbox{hadr})$ and $I_{AA}(\gamma+\mbox{jet})$.}

\label{away_yield}

\end{minipage}
\end{figure}

Figure \ref{icp_near} shows the ratio of associated near-side yield
per trigger particle in central and peripheral collisions as a
function of centrality: $I_{CP}^{near} = n_{near}(\mbox{central}) /
n_{near}(\mbox{peripheral})$.  The near-side yield for neutral BEMC
trigger clusters decreases by about 40\% with centrality for two
different $p_{T}^{assoc}$ bins (solid symbols). This feature is not
visible in correlations between charged particles with similar cuts
for trigger and associated particles (open symbols) \cite{dan/qm05},
which are consistent with an unmodified near-side correlation.  The
decrease of the associated yield in correlations between neutral and
charged particles can be attributed to single particle high-$p_T$
suppression, which reduces the pion yield per binary collision, but
does not affect prompt photon production
\cite{phenix/pi0,phenix/direct_photons}.  The result is a lower
fraction of $\pi^0$ triggers $N_{\pi^0}/(N_{\pi^0}+N_\gamma)$, and a
proportionally lower associated near-side yield per unidentified
trigger particle, if a constant yield per $\pi^0$ trigger is assumed.
This decrease of the $\pi^0$ fraction can be described by using the
$\pi^0$ fraction from a Pythia simulation as reference, scaling
$N_{\pi^0}$ with $R_{AA}$ from \cite{phenix/pi0}, and normalizing to
the most peripheral data point to allow comparison with
$I_{CP}^{near}$ (grey band). The data is consistent with the expected
decrease of the fraction of $\pi^0$ triggers from approximately 80\%
in peripheral to about half of all trigger particles in the most
central collisions.

Figure \ref{away_yield} shows the away-side yield per neutral trigger
particle with $E_{T}=10-15\,\mbox{GeV}$ as a function of centrality. The
away-side correlations have both $\gamma+$jet and dijet contributions.
We try to describe the away-side yield by adding the expected
$\gamma+$jet and dijet yields from a Pythia simulation after applying
two separate suppression factors $I_{AA}(\gamma+\mbox{jet})$ and
$I_{AA}(\mbox{hadr})$. The observed suppression is consistent with
both parameters being similar to the nuclear modification factor:
$I_{AA}(\gamma+\mbox{jet}) \approx I_{AA}(\mbox{hadr}) \approx
R_{AA}$. A separation of $\gamma+$jet and dijet correlations is
necessary to draw final conclusions.  The measurement of either
suppression factor is sufficient to determine both, as the combined
$I_{AA}$ is already known.

\begin{figure}
\hfil
\epsfig{file=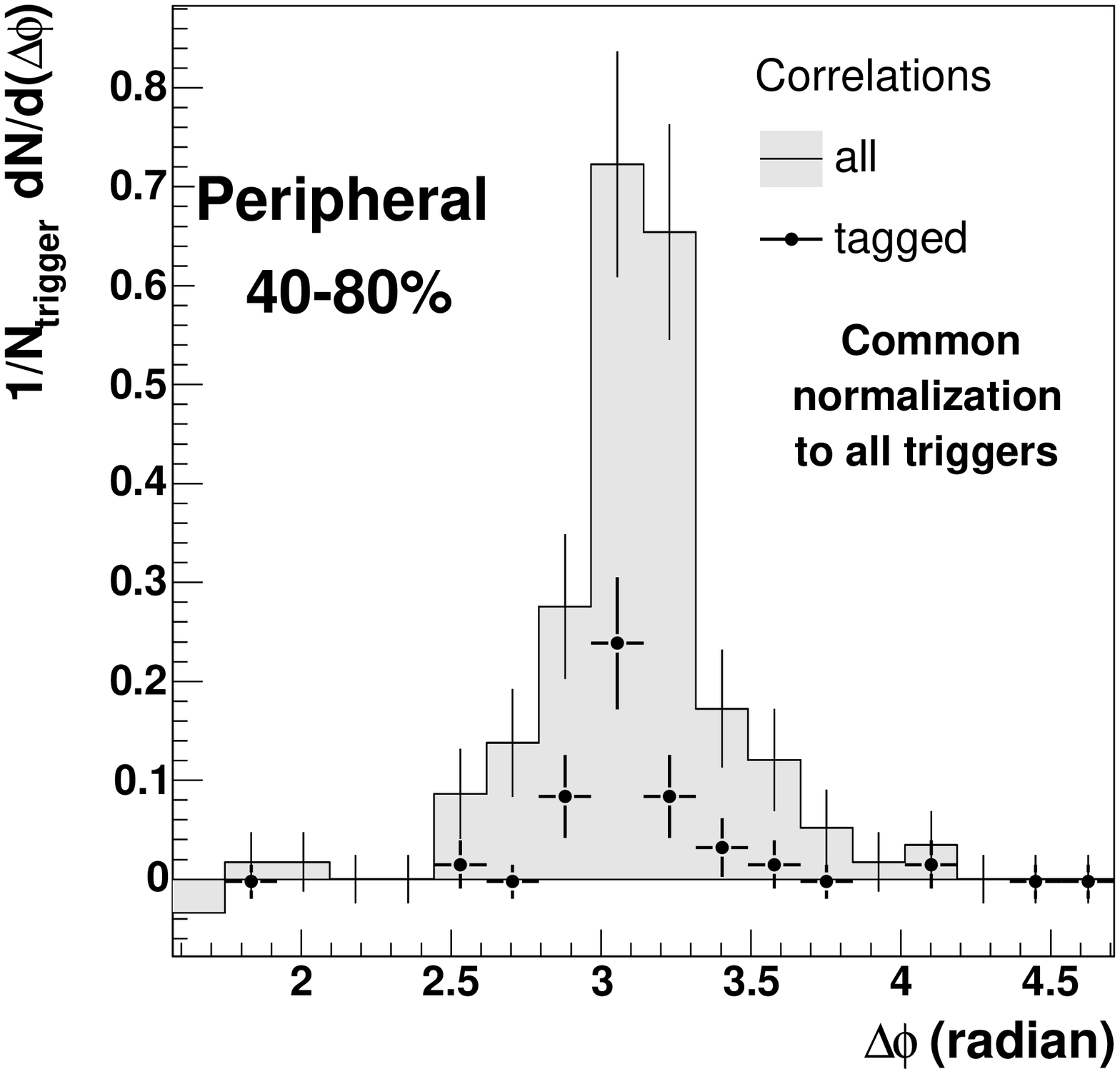, width=0.33\textwidth}
\hfil
\epsfig{file=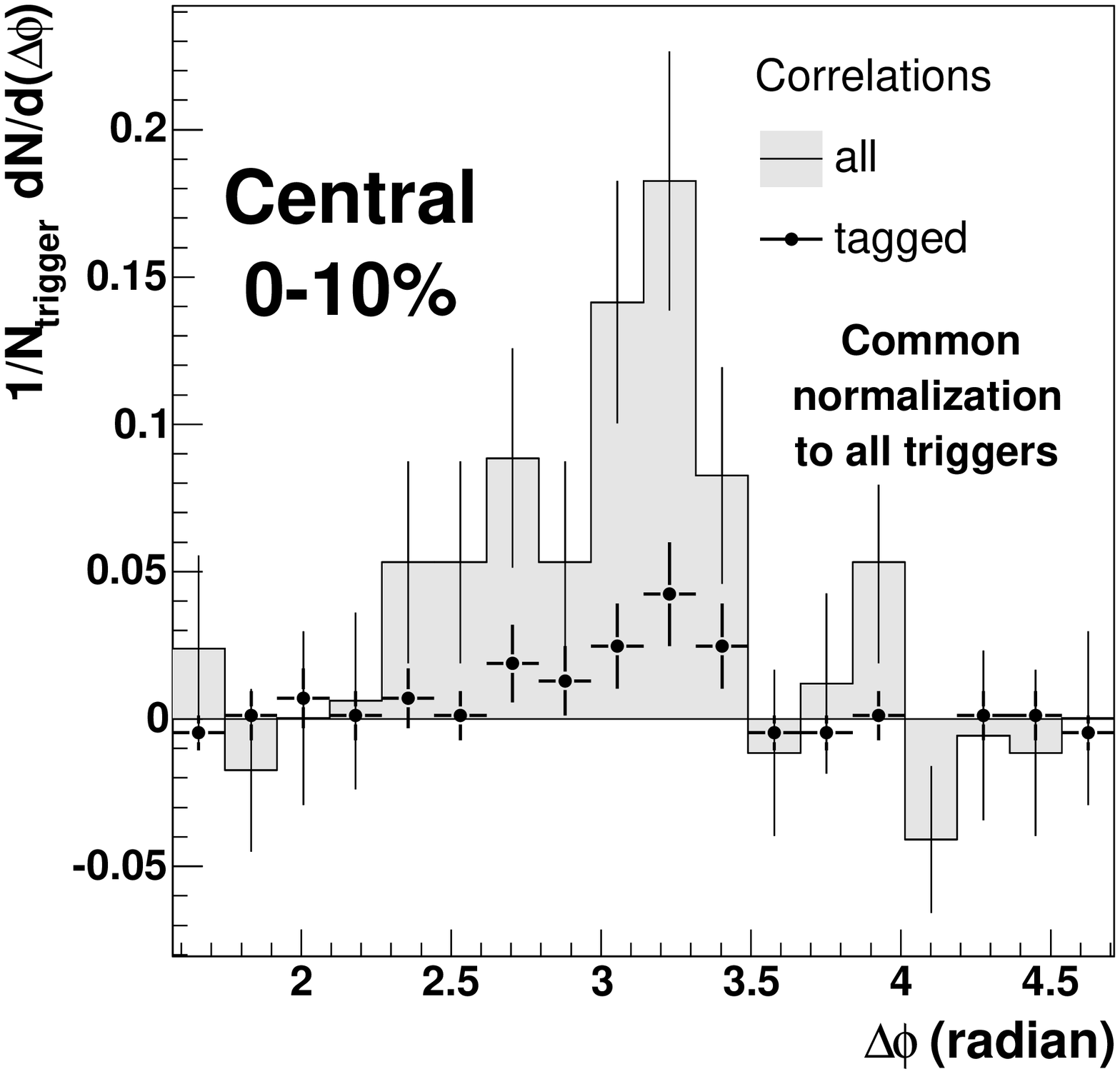, width=0.33\textwidth}
\hfil

\caption{Inclusive (shaded) and hadron-tagged (solid markers)
  away-side $\Delta\phi$ correlations with $E_{T}^{trigger} >
  10\,\mbox{GeV}$ and $3\,\mbox{GeV} < p_{T}^{assoc} <
  E_{T}^{trigger}$ for peripheral and central events. The
  hadron-tagged correlation has not been corrected for tagging
  efficiency.}

\label{htag}

\end{figure}

The absence of a near-side correlation in $\gamma+$jet events can be
used to tag dijet events by requiring a charged track with
$p_{T}>2.5\,\mbox{GeV}$ in a search cone around the cluster with a
radius of 0.15 in the $\eta$-$\phi$-plane. The cone size and $p_{T}$
cut have been chosen for high purity due to low random background, and
high efficiency, which is however limited by the near-side yield per
$\pi^0$ trigger. Figure \ref{htag} shows the away-side correlation for
these hadron-tagged and all, i.e.  tagged as well as untagged,
triggers, with a common normalization to the total number of neutral
trigger clusters, for two centrality bins.  The tagged correlation
function is not corrected for the tagging efficiency.  For both
centralities, about 20\% of all correlated back-to-back pairs have a
tagged trigger, indicating that dijets contribute substantially to
away-side correlations not only in peripheral, but also in central
events. A quantitative statement about the relative contributions from
$\gamma+$jet and dijets in central collisions will require a precise
determination of the tagging efficiency from simulations or peripheral
events, where dijet correlations dominate.

\section{CONCLUSIONS}

The dataset from the long Au+Au run in 2004 allows the analysis of
azimuthal correlations between photonic trigger particles with
energies above 10\,GeV and associated particles above 4\,GeV. A
decrease of the near-side yield indicates a large fraction of triggers
from $\gamma+$jet events (close to 50\%). The centrality dependence of
the away-side yield is consistent with similar values for the
away-side suppression $I_{AA}$ for dijet and $\gamma+$jet correlations
and the nuclear modification factor $R_{AA}$. The separation of dijet
and $\gamma+$jet contributions requires further analyses with better
identification of these two sources.

\end{document}